\documentclass[floatfix,5p]{elsarticle}

\usepackage{bm}
\usepackage{braket}
\usepackage{amssymb}
\usepackage{graphicx}
\usepackage{dcolumn}
\usepackage{amsmath}
\usepackage{amsfonts}
\usepackage{xcolor}
\usepackage{hyperref}

\hypersetup{
    colorlinks,
    linkcolor={red},
    citecolor={blue},
    urlcolor={green}
}

\biboptions{numbers,sort&compress}
\usepackage{mathtools}
\mathtoolsset{showonlyrefs}

\begin{document}
\title{Investigating the link between proton reaction cross sections and the quenching of proton spectroscopic factors in $^{48}$Ca}
\author[1,2]{M. C. Atkinson\corref{cor1}}
\address[1]{Department of Physics, Washington University in St. Louis, MO 63130 USA}
\address[2]{Theory Group, TRIUMF, BC V6T 2A3, Canada}
\ead{matkinson@triumf.ca}
\author[1]{W. H. Dickhoff} 
\cortext[cor1]{Corresponding author}

\date{\today}


\begin{abstract}
The nucleon self-energies of $^{40}$Ca and $^{48}$Ca are determined using a nonlocal dispersive optical model (DOM). By enforcing the dispersion relation connecting the real and imaginary part of the
self-energy, scattering and structure data are used to constrain these self-energies. The ability to calculate both bound and scattering states simultaneously puts these
self-energies in a unique position to consistently describe exclusive knockout reactions such as $(e,e'p)$.  The present  analysis reveals the importance of high-energy proton reaction cross-section data in
constraining spectroscopic factors required for the description of the $(e,e'p)$ cross sections. In particular, it is imperative that high-energy proton reaction cross-section data are measured for $^{48}$Ca in the near future so that the
quenching of the spectroscopic factors in the $^{48}$Ca$(e,e'p)^{47}$K reaction can be unambiguously constrained using the DOM. 
Measurements of proton reaction cross sections in inverse kinematics employing rare isotope beams with large neutron excess will provide corresponding constraints on proton spectroscopic factors for exotic nuclei.
Moreover, DOM generated spectral functions indicate that the quenching of
spectroscopic factors compared to $^{40}$Ca is not only due to long-range correlation, but also partly due to the increase in high-momentum protons in $^{48}$Ca on account of the strong neutron-proton interaction.
Single-particle momentum distributions of protons and neutrons in $^{48}$Ca calculated from these spectral functions confirm that neutron excess causes a higher fraction of
high-momentum protons than neutrons.  
\end{abstract}

\maketitle
\section{Introduction}
\label{sec:intro}
Independent particle models (IPMs) provide a simplified picture of the nucleus in which correlations are neglected and all orbitals are 100\% filled up to the Fermi level according to the Pauli principle and
those above it are empty.  However, due to residual interactions there is depletion of orbitals below the Fermi energy and filling of those above it.  The best tool to study this experimentally is the
$(e,e'p)$ reaction~\cite{Kramer:1989,Denherder:1988,Peter90,Lex90,Ingo91,Lapikas93,Pandharipande97}.  At sufficiently high electron energy and momentum transfer, the proton can be knocked out with enough
energy such that a description within the distorted-wave impulse approximation (DWIA) can be expected to be applicable, so that depletion (and also filling) of orbits can be studied~\cite{Denherder:1988,Kramer:1989}.
In the typical application of the DWIA to the $(e,e'p)$ reaction, a fully occupied IPM proton wave function is used which then requires a scaling factor of about 0.6-0.7 to describe the
overall magnitude of the data~\cite{Lapikas93}.  This scaling factor, usually referred to as the (reduced) spectroscopic factor, corresponds to the normalization of the overlap function between the target
ground state and low-lying single-hole states. Furthermore, the data show that additional removal strength with essentially the same overlap function is located at nearby energies, providing clear evidence of
the fragmentation of the single-particle strength~\cite{Kramer:1989,Kramer90}. 

This depletion of orbitals is closely linked with elastic scattering observables. Depletion becomes inevitable with the inclusion of a complex absorptive potential to account for inelastic processes in
the description of elastic scattering. A non-zero imaginary component of the optical potential at positive energies pulls strength away from the IPM orbitals. The reaction (total inelastic) cross section is the most sensitive to the
imaginary part of the optical potential, so it largely determines the depletion of these orbitals. 
In this way, the spectroscopic factors of orbitals are closely linked with the reaction cross section.
Thus, a proper description of $(e,e'p)$ data requires an optical potential that reproduces proton reaction cross-section data.
In Ref.~\cite{Atkinson:2018}, a nonlocal dispersive optical model (DOM) which simultaneously describes
both bound and scattering states was used to consistently provide all ingredients, including spectroscopic factors, for an accurate DWIA description of $^{40}$Ca$(e,e'p)^{39}$K data. 

A systematic study in Ref.~\cite{Gade:2014} summarized results for reduction factors obtained from nucleon-knockout reactions for a wide variety of nuclei.
The analysis employed results from shell-model calculations demonstrating that the removal of minority nucleons from nuclei with larger asymmetry leads to proportionally quenched reduction factors while nucleons of the majority species are less quenched.
This is not consistent with corresponding results of transfer reactions reviewed in Ref.~\cite{Dickhoff:2019} or the single-nucleon removal experiments recently reported in Refs.~\cite{Altar:2018,Kawase:2018}. 
At this time no clear consensus has been reached on this intriguing difference.
To investigate this discrepancy, a consistent DWIA
analysis of $^{48}$Ca$(e,e'p)^{47}$K is performed using a nonlocal DOM description similar to the one reported in Ref.~\cite{Atkinson:2018} for $^{40}$Ca. Comparing the DOM calculated spectroscopic factors of $^{48}$Ca and $^{40}$Ca will provide more information on the quenching of proton spectroscopic factors when neutrons are added.

The theoretical interpretation of the Nikhef $(e,e'p)$ results, reviewed in Refs.~\cite{Pandharipande97,Dickhoff04}, has mainly been concerned with the explanation of the reduction in the spectroscopic
strength to 60-70\% of the IPM value. While the main reduction of strength can be attributed to long-range correlations (LRC) which are manifest in the reaction cross section at lower energy, it has been well documented that
additional short-range and tensor correlations (SRC) are responsible for a 10-15\% depletion of the IPM value~\cite{Dickhoff04}.
These SRCs give rise to high-momentum nucleon pairs which have been measured with inclusive $(e,e')$ inelastic scattering by the Continuous Electron Beam Accelerator Facility (CEBAF) Large Acceptance
Spectrometer (CLAS) collaboration at Jefferson Lab in $^3$He, $^4$He, $^{12}$C, and $^{56}$Fe~\cite{CLAS:2006}. 
Asymmetric nuclear-matter calculations for various realistic interactions have documented the importance of the tensor force in generating a larger depletion of the proton Fermi sea compared to the neutron one when protons are in the minority, thus generating relatively more high-momentum protons than neutrons~\cite{Rios:2009,Rios:2014}.
Realistic many-body calculations of low-$A$ nuclei using variational Monte Carlo (VMC)
techniques also reveal that the majority of this high-momentum content comes from the tensor force in the nucleon-nucleon interaction~\cite{Wiringa:2014}. This non-negligible fraction of high-momentum nucleons is
further proof that there are correlations beyond the mean-field in nuclei. This high-momentum content can be calculated in the DOM framework, which provides another means of investigating the quenching of the
spectroscopic factor and many-body correlations in $^{40}$Ca and $^{48}$Ca.


\section{Analysis of $^{48}$Ca$(e,e'p)^{47}$K reaction employing the nonlocal DOM}
\label{sec:theory}

\begin{figure}[h]
   \begin{minipage}{\columnwidth}
      \makebox[\columnwidth]{
         \includegraphics[scale=1.0]{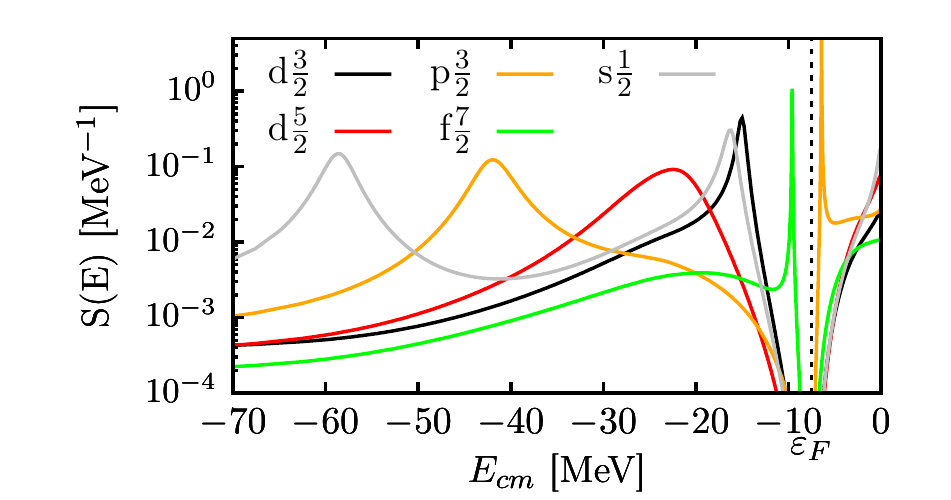}
      }
   \end{minipage}
   \caption{Neutron spectral functions of a representative set of $\ell j$ shells in $^{48}$Ca. The particle states are distinguished from the hole states by the dotted line representing the Fermi energy.}
   \label{fig:spectral_n}
\end{figure}

\begin{figure}[h]
   \begin{minipage}{\columnwidth}
      \makebox[\columnwidth]{
         \includegraphics[scale=1.0]{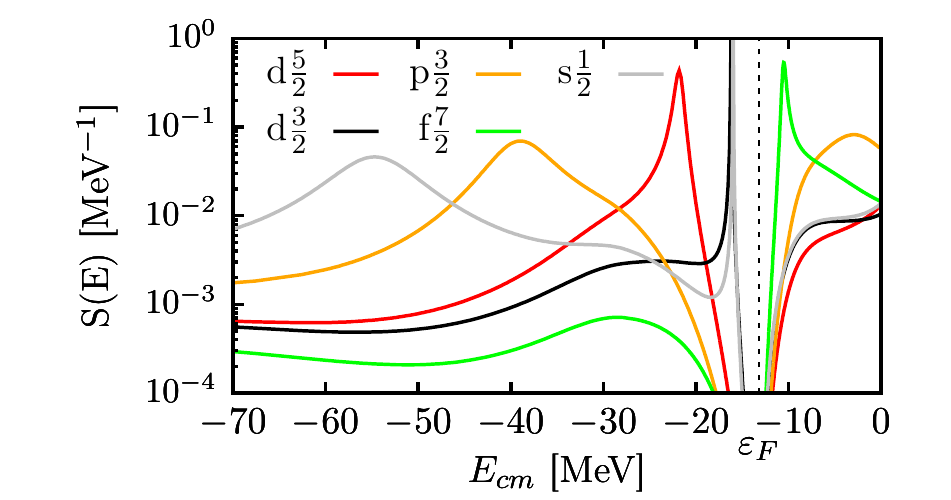}
      }
   \end{minipage}
   \caption{Proton spectral functions of a representative set of $\ell j$ shells in $^{48}$Ca. The particle states are differentiated from the hole states by the dotted line representing the Fermi energy.}
   \label{fig:spectral_p}
\end{figure}

The nonlocal dispersive-optical-model (DOM) uses both bound and scattering data to constrain the nucleon self-energy  $\Sigma_{\ell j}$ for a given nucleus.
This self-energy is a complex and nonlocal potential that unites the nuclear structure and reaction domains~\cite{Mahaux91,Mahzoon:2014}.
 The DOM was originally developed by Mahaux and Sartor~\cite{Mahaux91}, employing local real and imaginary potentials connected through dispersion relations. However, only
with the introduction of nonlocality can realistic self-energies be obtained~\cite{Mahzoon:2014,Dickhoff:2017}. 
The Dyson equation then determines the single-particle propagator or Green's function $G_{\ell j}(r,r';E)$ from which bound-state and scattering observables can be deduced. 
The hole spectral density for energies below the Fermi energy $\varepsilon_F$ is obtained from the single-particle propagator in the following way, 
\begin{equation}
   S^h_{\ell j}(\alpha,\beta;E) = \frac{1}{\pi}\textrm{Im}\ G_{\ell j}(\alpha,\beta;E) .
   \label{eq:spec}
\end{equation}
The diagonal element of Eq.~\eqref{eq:spec} is known as the (hole) spectral function identifying the probability density for the removal of a single-particle state with quantum numbers $\alpha \ell j$ at
energy $E$.
The spectral strength for a given $\ell j$ combination can be found by summing (integrating) the spectral function according to
\begin{equation}
   S_{\ell j}(E) = \sum_\alpha S_{\ell j}(\alpha,\alpha;E) .
   \label{eq:strength}
\end{equation}
The spectral strength $S_{\ell j}(E)$ is the contribution at energy $E$ to the
occupation from all orbitals with $\ell j$. It reveals that the strength for these
shells is fragmented, rather than concentrated at the independent-particle model
(IPM) energy levels.  Figure~\ref{fig:spectral_n} shows the spectral strength for a
representative set of neutron shells in $^{48}$Ca that would be considered bound in
the IPM. The peaks in Fig.~\ref{fig:spectral_n} correspond to the binding energies of
the appropriate IPM orbitals. For example, the p$\frac{3}{2}$ spectral function in
Fig.~\ref{fig:spectral_n} has two peaks, one below $\varepsilon_F$ corresponding to
the 0p$\frac{3}{2}$ quasihole state, and one above $\varepsilon_F$ corresponding to
the 1p$\frac{3}{2}$ quasiparticle state.  Comparing the neutron spectral functions
in Fig.~\ref{fig:spectral_n} with the proton spectral functions in
Fig.~\ref{fig:spectral_p} reveals that the proton peaks are broader at a similar distance from the corresponding Fermi energy than those of the neutrons. 
The larger broadening of these peaks is a consequence of the protons being
more correlated than the neutrons as determined by the fit to all relevant
experimental data generating larger absorptive potentials for protons than neutrons at all energies. 

The occupation of specific orbitals characterized by $n$ with wave functions that
are normalized to 1 can be obtained from Eq.~\eqref{eq:spec} by folding in the corresponding wave functions~\cite{Dussan:2014},
\begin{equation}
   S^{n-}_{\ell j}(E) = \sum_{\alpha,\beta}[\phi^n_{\ell j}(\alpha)]^*S^h_{\ell j}(\alpha,\beta;E)\phi^n_{\ell j}(\beta) .
   \label{eq:qh_strength}
\end{equation}
Note that this representation of the spectral strength involves off-diagonal elements of the propagator.
The wave functions used in Eq.~\eqref{eq:qh_strength} are the solutions of the Dyson equation that correspond to discrete bound states with one proton/neutron removed.
Such quasihole wave functions can be obtained from the nonlocal Schr\"{o}dinger-like equation disregarding the imaginary part
\begin{align}
   \sum_\gamma\bra{\alpha}T_{\ell} + \textrm{Re}\ \Sigma^*_{\ell j}(\varepsilon_n^-)\ket{\gamma}\psi_{\ell j}^n(\gamma) = \varepsilon_n^-\psi_{\ell j}^n(\alpha),
   \label{eq:schrodinger}
\end{align}
where $\bra{\alpha}T_\ell\ket{\gamma}$ is the kinetic-energy matrix element, including the centrifugal term.
These wave functions correspond to overlap functions
\begin{equation}
   \psi^n_{\ell j}(\alpha) = \bra{\Psi_n^{A-1}}a_{\alpha \ell j}\ket{\Psi_0^A}, \qquad \varepsilon_n^- = E_0^A - E_n^{A-1}.
   \label{eq:wavefunction}
\end{equation}
Such discrete solutions to Eq.~\eqref{eq:wavefunction} exist where there is no imaginary part of the self-energy, so near the Fermi energy. 
The normalization for these wave functions is the spectroscopic factor, which is given by~\cite{Exposed!}
\begin{equation}
   \mathcal{Z}^n_{\ell j} = \bigg(1 - \frac{\partial\Sigma_{\ell j}^*(\alpha_{qh},\alpha_{qh};E)}{\partial E}\bigg|_{\varepsilon_n^-}\bigg)^{-1},
   \label{eq:sf}
\end{equation}
where $\alpha_{qh}$ corresponds to the quasihole state that solves Eq.~\eqref{eq:schrodinger}. This corresponds to the spectral strength at the quasihole energy $\varepsilon_n^-$, represented by a delta
function. The quasihole peaks in Fig.~\ref{fig:spectral_p} get narrower as the levels approach $\varepsilon_F$, which is a consequence of the imaginary part of the irreducible self-energy decreasing when
approaching $\varepsilon_F$. In fact, the last mostly occupied proton level in Fig.~\ref{fig:spectral_p} (1s$\frac{1}{2}$) has a spectral function that is essentially a delta function peaked at its energy
level, where the imaginary part of the self-energy vanishes.  For these orbitals, the strength of the spectral function at the peak corresponds to the spectroscopic factor in Eq.~\eqref{eq:sf}. Note that
because of the presence of imaginary parts of the self-energy at other energies, there is also strength located there, thus the spectroscopic factor will be less than one and also less than the occupation
probability.

\begin{figure}[h]
   \begin{minipage}{\columnwidth}
      \makebox[\columnwidth]{
         \includegraphics[scale=1.0]{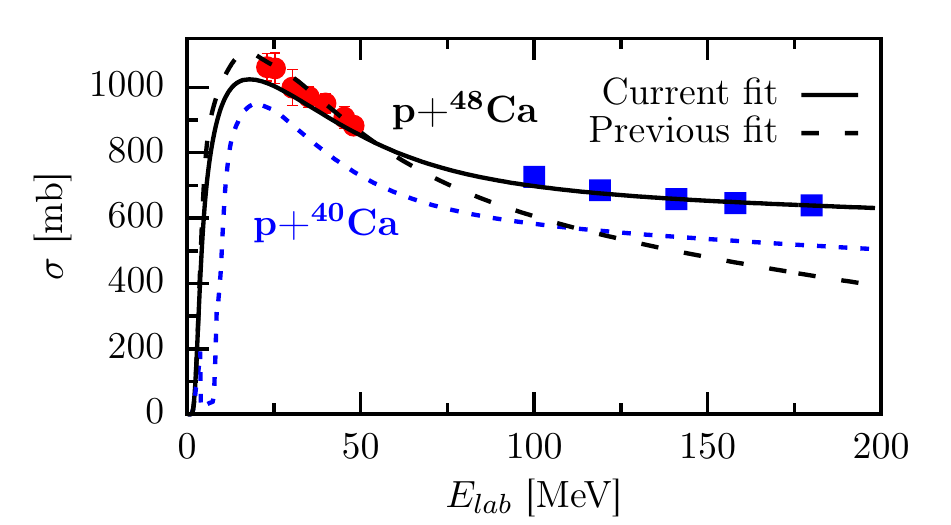}
      }
   \end{minipage}
   \caption[The proton reaction cross section for $^{48}$Ca.]
   {Proton reaction cross sections for $^{48}$Ca and $^{40}$Ca. The solid line represents the current $^{48}$Ca fit while the dashed line depicts the previous $^{48}$Ca fit~\cite{Mahzoon:2017}. The dotted
   line represents the $^{40}$Ca fit from Ref.~\cite{Atkinson:2018}. The circular points are the same experimental data used in Ref.~\cite{Mueller:2011} and were included in the previous fit. The square points
   are extrapolated from the $\sigma_\text{react}^{40}(E)$ experimental data points at the corresponding energies. The extrapolation is explained in the main text.}
   \label{fig:react48}
\end{figure}

Previously, a fit of $^{48}$Ca was published in Ref.~\cite{Mahzoon:2017}, quoting a neutron skin of $\Delta r_{np} = 0.249\pm0.023$~fm. However, just as in the case of $^{40}$Ca in
Refs.~\cite{Mahzoon:2014,Atkinson:2018}, the proton reaction cross section is underestimated at 200 MeV. While there are no experimental data for $^{48}$Ca at these energies, there is a data point at 700 MeV of the
proton reaction cross section of $^{40}$Ca and $^{48}$Ca~\cite{Anderson700}. Comparing the available data for $\sigma_\text{react}^{40}(E)$ at 200 MeV and 700 MeV reveals that the reaction cross section essentially stays flat
between these energies. It is reasonable to expect that $\sigma_\text{react}^{48}(E)$ assumes the same shape as $\sigma_\text{react}^{40}(E)$
at high energies. Thus, data points are extrapolated from the $^{40}$Ca experimental data at energies above 100 MeV by applying the ratio that is seen in the 700 MeV data for
$\sigma_\text{react}^{48}(E)/\sigma_\text{react}^{40}(E)$, see
Table~\ref{table:high_react}. The extrapolated points are shown as blue squares in Fig.~\ref{fig:react48} while the updated fit is represented with the solid curve. 
The remainder of the fit did not change significantly from
Ref.~\cite{Mahzoon:2017}. The parameterization of the $^{48}$Ca self-energy as well as the updated parameters are presented in the supplementary material.

\begin{table}[h]
   \caption[Experimental proton reaction cross-section data at 700 MeV]
   {Experimental proton reaction cross-section data at 700 MeV taken from
      Ref.~\cite{Anderson700}. 
   }
   \label{table:high_react}
   \resizebox{\columnwidth}{!}{
      \begin{tabular}{c c c c }
         \hline
         \hline
         Nucleus  & $^{40}$Ca & $^{48}$Ca & $^{48}$Ca/$^{40}$Ca \\
         \hline
         $\sigma_{react}(E)$  & $614\pm38$~mb  & $736\pm46$~mb & 1.19 \\
         \hline
         \hline
      \end{tabular}
   }
\end{table}

\begin{figure}[h]
   \begin{minipage}{\columnwidth}
      \makebox[\columnwidth]{
         \includegraphics[scale=1.0]{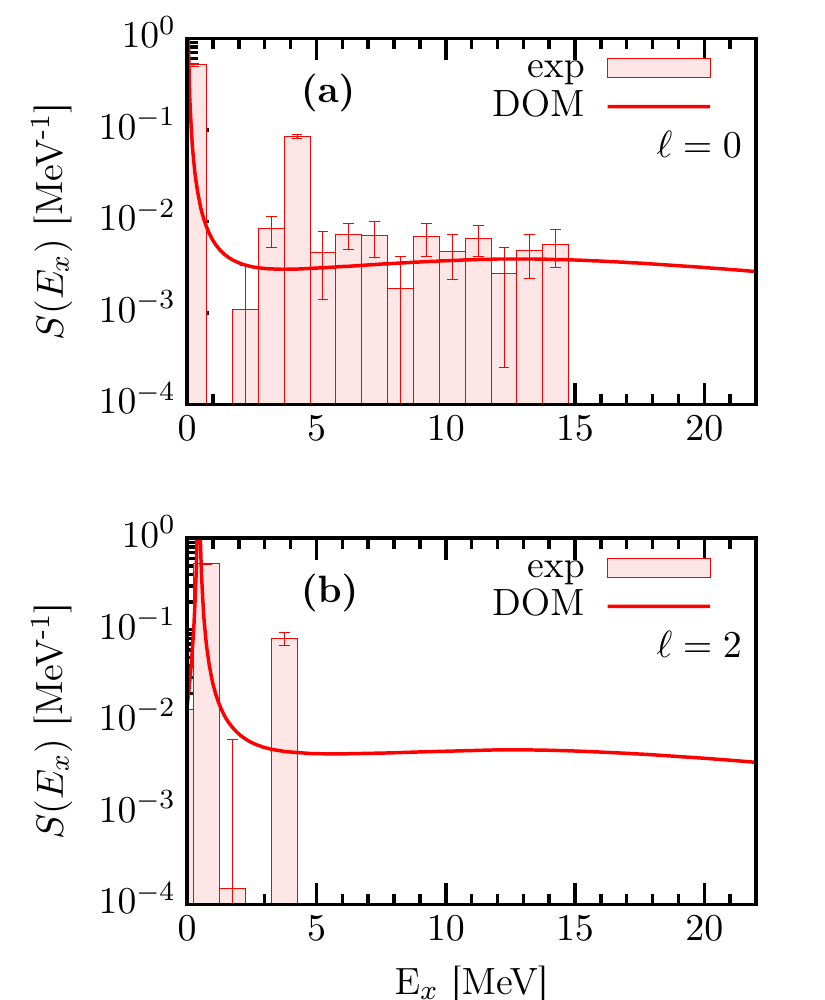}
      }
   \end{minipage}
   \caption[Spectral strength as a function of excitation energy in $^{48}$Ca]{
      Spectral strength as a function of excitation energy in $^{48}$Ca. The solid lines are DOM spectral functions for (a) the $1\textrm{s}\frac{1}{2}$ and (b) the $0\textrm{d}\frac{3}{2}$ proton orbitals.
      The histograms are the excitation energy spectra in $^{39}$K  extracted from the $^{48}$Ca$(e,e'p)$$^{47}$K experiment~\cite{Kramer:2001,Kramer90}.  The  peaks in the DOM curves and experimental data
      correspond to the quasihole energies of the protons in $^{40}$Ca. The experimental spectrum in (b) is the isolated $0$d$\frac{3}{2}$ orbital.
   } 
   \label{fig:spectral_48}
\end{figure}

To analyze the proton spectroscopic factors, the \\ $^{48}$Ca$(e,e'p)^{47}$K cross section is calculated using the DWIA following the same procedure detailed in
Ref.~\cite{Atkinson:2018} for $^{40}$Ca. In the DWIA, the $(e,e'p)$ cross section is calculated using a distorted wave to represent the outgoing proton, a proton bound-state wave function (BSWF) representing the
struck proton, and the normalization of the BSWF corresponding to the spectroscopic factor. All of these quantities are directly provided by the DOM self-energy.
The experimental data of the $^{48}$Ca$(e,e'p)^{47}$K reaction were obtained in parallel kinematics for outgoing proton kinetic energies of $T_p = 100$~MeV at Nikhef
and previously published in Ref.~\cite{Kramer:2001}. Just as in Ref.~\cite{Atkinson:2018}, the DOM spectroscopic factors need to be renormalized by incorporating the observed experimental fragmentation of the strength near the Fermi energy that is not yet included in the DOM self-energy. The experimental strength distributions for the $\ell=0$ and the $\ell=2$ excitations of $^{47}$K are shown in Fig.~\ref{fig:spectral_48}, overlaid with the corresponding DOM spectral
functions calculated from Eq.~\eqref{eq:qh_strength}.  Analogously to the $^{40}$Ca calculation, the distributions in Fig.~\ref{fig:spectral_48} are used to renormalize the DOM spectroscopic factors in the
following way, 
\begin{equation}
   \frac{\mathcal{Z}_F^{\text{DOM}}}{\int dE\ S^{\text{DOM}}(E)} = \frac{\mathcal{Z}_F^{\text{exp}}}{\int dE\ S^{\text{exp}}(E)}.
   \label{eq:ratio}
\end{equation}
This scaling results in a reduction from 0.64 to 0.55 for the $1$s$\frac{1}{2}$ orbital and from 0.60 to 0.58 for the $0$d$\frac{3}{2}$ orbital. These values are in good agreement with
originally published spectroscopic factors~\cite{Kramer:2001}, as seen in Table~\ref{table:sf_48}. The uncertainties in the values of the spectroscopic factors were determined using the same bootstrap method discussed in the previous DOM analysis of $^{40}$Ca$(e,e'p)^{39}$K~\cite{Atkinson:2018}. 

\begin{table}[h]
   \caption[Comparison of spectroscopic factors in $^{48}$Ca]{Comparison of spectroscopic factors in $^{48}$Ca deduced from the previous analysis~\cite{Kramer:2001} using the Schwandt optical potential~\cite{Schwandt:1982} to the normalization of the corresponding overlap functions obtained in the present analysis from the DOM including an error estimate as described in the text.}
   \vspace{0.5cm}
   \begin{minipage}{\columnwidth}
      \makebox[\columnwidth]{
         \begin{tabular}{ c c c } 
            \hline
            \hline
            $\mathcal{Z}$ & $0\textrm{d}\frac{3}{2}$ & $1\textrm{s}\frac{1}{2}$\\
            \hline
            Ref.~\cite{Kramer:2001} & $0.57 \pm 0.04$ & $0.54 \pm 0.04$\\
            \hline
            DOM & $0.58 \pm 0.03$ & $0.55 \pm 0.03$ \\
            \hline
            \hline
         \end{tabular}
      }
   \end{minipage}
   \label{table:sf_48} 
\end{table}

\begin{figure}[h]
   \begin{minipage}{\columnwidth}
      \makebox[\columnwidth]{
         \includegraphics[scale=1.0]{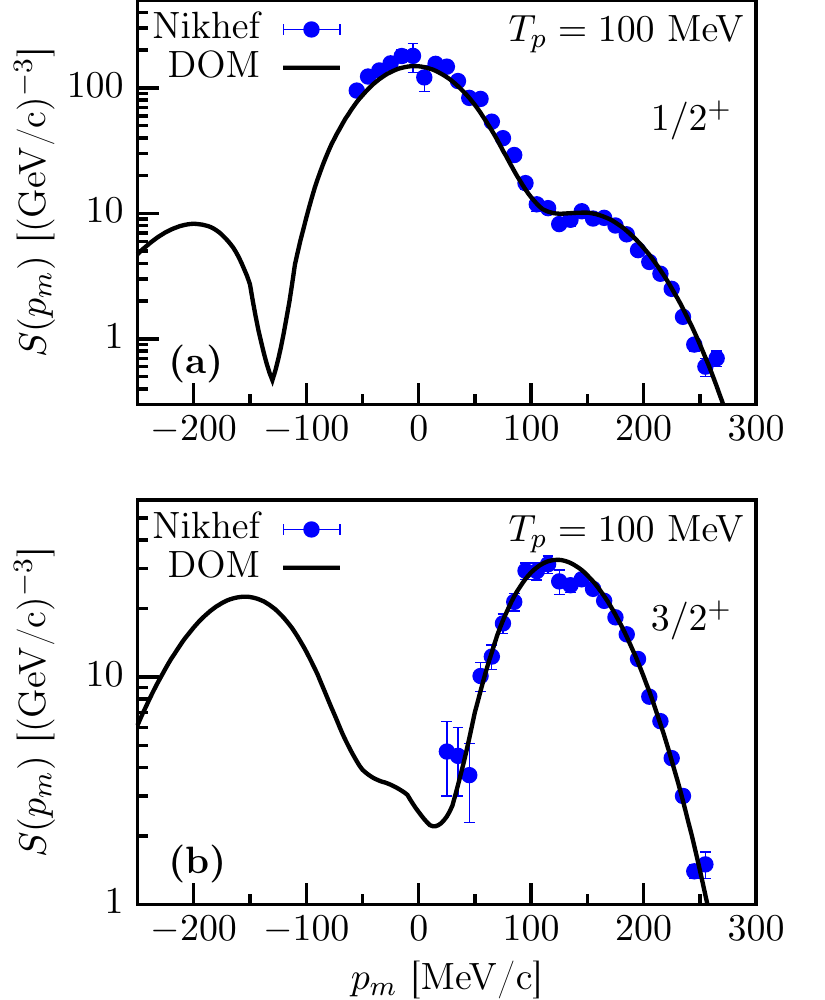}
      }
   \end{minipage}
   \caption[$^{48}$Ca$(e,e'p)$$^{47}$K spectral functions in parallel kinematics at an outgoing proton kinetic energy of 100 MeV]{
      $^{48}$Ca$(e,e'p)$$^{47}$K spectral functions in parallel kinematics at
      an outgoing proton kinetic energy of 100 MeV. The solid line is the
      calculation using the DOM ingredients while the points are from the
      experiment detailed in Ref.~\cite{Kramer:2001}.  (a) Distribution for the
      removal of the $1\textrm{s}\frac{1}{2}$ proton. The curve contains the
      DWIA for the $1/2^+$ ground state using the DOM generated spectroscopic
      factor of 0.55 (renormalized using Eq.~\eqref{eq:ratio})  (b)
      Distribution for the removal of the $0\textrm{d}\frac{3}{2}$ with a DOM
      generated spectroscopic factor of 0.58 (renormalized using
      Eq.~\eqref{eq:ratio}) for the $3/2^+$ excited state at 0.36 MeV.
}
   \label{fig:eep_48}
\end{figure}

Using the resulting renormalized spectroscopic factors produces the momentum distributions shown in Fig.~\ref{fig:eep_48}.  Thus, the smaller spectroscopic factors in $^{48}$Ca are consistent with the
experimental cross sections of the \\ $^{48}$Ca$(e,e'p)^{47}$K reaction. The comparison of $\mathcal{Z}_{48}$ and $\mathcal{Z}_{40}$ in Table~\ref{table:sf_comp} reveals that both orbitals experience a
reduction. This indicates that strength from the spectroscopic factors is pulled to the continuum in $S(E)$ when eight neutrons are added to $^{40}$Ca. 
Thus, the stronger coupling to surface excitations in $^{48}$Ca, demonstrated by the larger proton reaction cross section when compared to $^{40}$Ca (see Fig.~\ref{fig:react48}), strongly contributes to the quenching of the proton
spectroscopic factor. It is important to note how crucial the extrapolated high-energy proton reaction cross-section data are in drawing
these conclusions. Without them, there is no constraint for the strength of the spectral function at large positive energies, which could result in no quenching of the spectroscopic factors of $^{48}$Ca due
to the sum rule that requires the strength to integrate to one when all energies are considered~\cite{Dussan:2014,Exposed!}.

\begin{table}[h]
   \caption{Comparison of DOM spectroscopic factors in $^{48}$Ca and $^{40}$Ca.
   These factors have not been renormalized and represent the aggregate strength near the Fermi energy.}
   \vspace{0.5cm}
   \begin{minipage}{\columnwidth}
      \makebox[\columnwidth]{
         \begin{tabular}{ c c c } 
            \hline
            \hline
            $\mathcal{Z}$ & $0\textrm{d}\frac{3}{2}$ & $1\textrm{s}\frac{1}{2}$\\
            \hline
            $^{40}$Ca & $0.71 \pm 0.04$ & $0.74 \pm 0.03$ \\
            \hline
            $^{48}$Ca & $0.60 \pm 0.03$ & $0.64 \pm 0.03$ \\
            \hline
            \hline
         \end{tabular}
      }
   \end{minipage}
   \label{table:sf_comp} 
\end{table}

In addition to the depletion of the spectroscopic factor due to LRC, strength is also pulled to continuum energies due to SRC.
It was stated earlier that a large portion of high-momentum content is caused by the tensor force in the nucleon-nucleon interaction. In particular, the tensor force preferentially acts on pairs of neutrons
and protons ($np$ pairs) with total spin $S=1$. This phenomenon is known as $np$ dominance~\cite{Hen:2017}, and is demonstrated by a factor of 20 difference between the number of observed $np$ SRC pairs and
the number of observed $pp$ and $nn$ SRC pairs in exclusive $(e,e'pp)$ and $(e,e'p)$ cross section measurements of $^{12}$C, $^{27}$Al, $^{56}$Fe, and $^{208}$Pb~\cite{Hen:2017}.
The dominance of $np$ SRC pairs would imply that the number of high-momentum protons observed in a nucleus is dependent on how many neutrons it contains. More specifically, one would expect that the
high-momentum content of protons would increase with neutron excess since there are more neutrons available to make $np$ SRC pairs. The CLAS collaboration confirmed this asymmetry dependence by measuring the
high-momentum content of protons and neutrons from $(e,e'p)$ and $(e,e'n)$ cross section measurements in $^{12}$C, $^{27}$Al, $^{56}$Fe, and $^{208}$Pb~\cite{Duer:2018}.

\begin{figure}[h]
   \begin{minipage}{\columnwidth}
      \makebox[\columnwidth]{
         \includegraphics[scale=1.0]{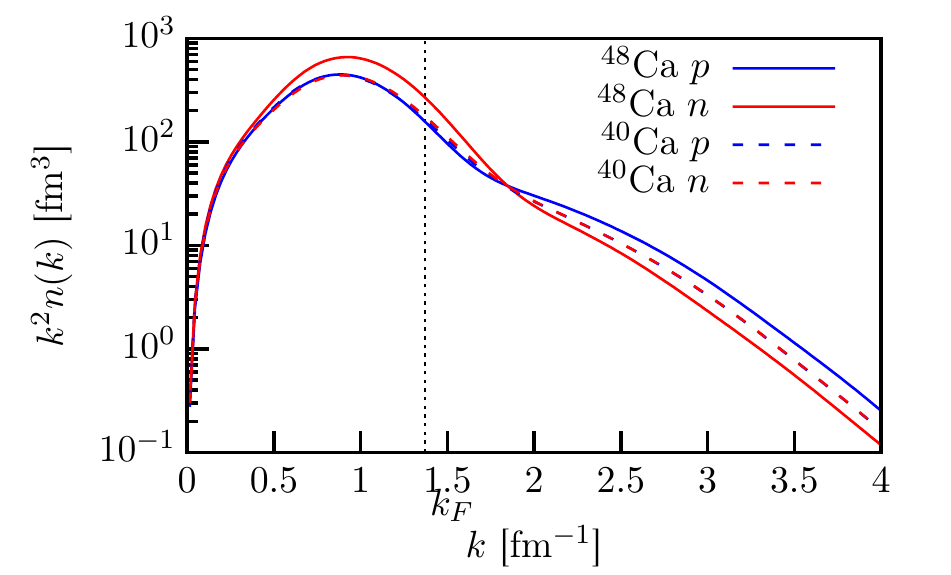}
      }
   \end{minipage}
   \caption[Comparison of DOM calculated momentum distribution of protons and
   neutrons in $^{48}$Ca and $^{40}$Ca.]
   {
      Comparison of DOM calculated momentum distributions of protons (blue) and
      neutrons (red) in $^{48}$Ca (solid) and $^{40}$Ca (dashed). The dotted line
      marks the value used for $k_F$. 
   }
   \label{fig:kcomp}
\end{figure} 

This effect can be studied by comparing the DOM generated momentum distributions
for $^{40}$Ca and $^{48}$Ca, since the only difference between them is the eight
additional neutrons in $^{48}$Ca mainly filling the 0f$\frac{7}{2}$ shell. The momentum
distributions for $^{40}$Ca and $^{48}$Ca are shown in Fig.~\ref{fig:kcomp}.
It is clear that the $^{48}$Ca proton momentum distribution (solid blue line) has
more high-momentum content than the $^{40}$Ca proton momentum distribution (dashed
blue line). Furthermore, since the number of protons does not change between
$^{40}$Ca and $^{48}$Ca, the added high-momentum content in the tail of $^{48}$Ca
is accounted for by a reduction of the distribution in the $k<k_F$ region. Turning now to the neutrons in
Fig.~\ref{fig:kcomp}, the $^{48}$Ca momentum distribution is larger in
magnitude than the $^{40}$Ca distribution for $k<k_F$. This is not
surprising since there are now eight more neutrons which are dominated by low-momentum content. The high-momentum content of the neutrons in $^{40}$Ca decreases from $14.7\%$ to $12.6\%$ when eight neutrons are added to form $^{48}$Ca while the high-momentum content of the protons increases from 14.0\% to 14.6\%.
The effects of the asymmetry of $^{48}$Ca on high-momentum content are evident in the fact that there are now significantly more high-momentum protons than neutrons. Both
the increase in proton high-momentum content and the decrease in neutron high-momentum content are qualitatively consistent with the CLAS
measurements of neutron-rich nuclei~\cite{Duer:2018} and support the $np$-dominance picture as predicted in Refs.~\cite{Rios:2009,Rios:2014}. Note that at this stage of the DOM development, no attempt has been made to quantitatively account for the CLAS observations.


Another manifestation of the more correlated protons can be seen in the spectral
functions of Figs.~\ref{fig:spectral_n} and~\ref{fig:spectral_p}. The broader peaks
of the proton spectral functions indicate that the protons are more correlated.
Furthermore, increased proton high-momentum content in $^{48}$Ca comes from generating
more strength in the continuum of the hole spectral function than in $^{40}$Ca. To
compare how strength is distributed over energy in $^{40}$Ca and $^{48}$Ca, the sum
over all $\ell j$ shells can be performed, 
\begin{equation}
   S(E) = \sum_{\ell j}^\infty (2j+1)S_{\ell j}(E),
   \label{eq:spectral_sum}
\end{equation}
where $S_{\ell j}(E)$ are defined in Eq.~\eqref{eq:strength}. The summed
spectral function of $^{48}$Ca has more strength than that of
$^{40}$Ca at large negative energies. In order to conserve proton number, an
increase in strength at continuum energies in $S(E)$ of $^{48}$Ca must be
compensated by a decrease in strength from energies close to the proton Fermi energy
in $^{48}$Ca.  In particular, this contributes to the quenching of the spectroscopic
factors of the $0$d$\frac{3}{2}$ and $1$s$\frac{1}{2}$ orbitals, before
renormalization (see Eq.~\eqref{eq:ratio}), in $^{48}$Ca from the values for
$^{40}$Ca as can be seen in Table~\ref{table:sf_comp}.  In this way, the
spectroscopic factor provides a link between the low-momentum knockout experiments
done at Nikhef and the high-momentum knockout experiments done at JLAB by the CLAS
collaboration.   

\section{Summary}
\label{sec:conclusions}

The DOM analysis of the $^{40,48}$Ca$(e,e'p)^{39,47}$K reactions demonstrates that the addition of eight neutrons to $^{40}$Ca leads to a quenching of the proton spectroscopic factors, in agreement with the trend observed in Ref.~\cite{Gade:2014} but with a reduced slope. Some form of quenching is inevitable if one accepts the $np$ dominance picture, since the added neutrons cause the protons to become more correlated. The increase in the
high-momentum content of protons in $^{48}$Ca is consistent with the $np$ dominance picture, hence it contributes to the quenching of the spectroscopic factors. Additionally, the increased proton reaction
cross section of $^{48}$Ca at all energies compared to $^{40}$Ca leads to more depletion, which also contributes to the observed quenching.  The proton reaction cross section plays a delicate role in determining the
spectroscopic factor.  While in the case of $^{48}$Ca the lack of proton reaction
cross-section data points at energies between $100$-$200$ MeV was compensated for by modifying the corresponding $^{40}$Ca data
points, precise measurements of the proton reaction cross sections at these energies are crucial in constraining spectroscopic factors.
Such measurements in inverse kinematics with rare isotopes can further help understand the behavior of spectroscopic factors away from the valley of stability.

\section*{Acknowledgements}
   This work was supported by the U.S. National Science Foundation under grants PHY-1613362 and PHY-1912643.

\bibliographystyle{elsarticle-test}
\bibliography{ca48-react}

\end{document}